# The Four Space-times Model of Reality

Giorgio Fontana

*University of Trento, I-38050 Povo, Italy*
*+39-0461-883906; giorgio.fontana@unitn.it*

**Abstract.** We live in a 3+1 space-time that is intended as a description of the universe with three space dimensions and one time dimension. Space-time dimensionality seems so natural that it is rarely criticized. Experiments and the highly successful relativistic theories teach us that there are four fundamental dimensions, among them is time that is treated as a special dimension. The specialty of time can be removed, leading to the concept that time is simply a function of four new fundamental dimensions, which have now identical properties, in combination with Lorentz invariance. A model is deduced in which a 4-space, characterized by four space-like coordinates, may host four "equivalent but orthogonal" space-times, each with three spatial coordinates and one temporal coordinate. Coordinates are shared; therefore the 4-space and the four space-times are all in one. Electromagnetic interaction is confined in each space-time and the role of the speed of light appears to be that of a barrier for the electromagnetic interaction. The motion of objects can be described by four-dimensional optics in the 4-space. Each of the four space-times may host a universe and, in agreement with recent observations, the proposed model can be directly applied to problems like the cosmological matter-antimatter asymmetry and dark-matter issues. Space travel may also benefit from the concepts presented.

## INTRODUCTION

One of the unsolved mysteries that science is facing and trying to solve is the origin of the observed matter-antimatter asymmetry (Dine, 2004), even if the simplest model of the origin of the universe is in favor of a perfect symmetry, observed in accelerator experiments.

The antimatter mystery is not alone. Observation of the local surroundings and of the far reaches of the universe with any possible optical instrument only shows the presence of void, electromagnetic radiation and matter, more precisely baryon matter (consisting of strongly interacting fermions like protons, electrons and neutrons).

Taking into account gravitational phenomena, a puzzling discovery has been made. Baryon matter is only 5% of the total energy density of the universe, less than 1/4 is some nonbaryonic matter, 3/4 is in a form of negative pressure, named cosmological constant (Peebles, 2003; Deffayet, 2002). Dark matter seems composed of a type of invisible objects of unknown size, that have mass but stay diluted and do not gravitationally accumulate to produce stars or planets. These objects do indeed accumulate at scales of the size of a galaxy, where their existence has been inferred; they may also induce the formation of the galaxies themselves. How can this happen? Being the nonbaryonic matter here within us, why we cannot see it, either directly, with a range of already built detectors or by any possible obscuration effect at the galactic scale? Why the exact nature of the nonbaryonic matter is not yet predicted with a model of particles that is, for some respects, quite accurate? Does it really exist in "our" local/visible universe? Can the understanding of the dark-matter problem be of any usefulness to correctly approach the cosmological constant problem or become the roots to quantum mechanics?

The new paradigm considered is that "visible" baryon matter is a phenomenon that take place in the space-time we live in, and "living" means "producing electromagnetic phenomena". On the contrary, dark matter populates one or more parallel universes that share one or more space-time dimensions with "our" local universe, which is embedded in "our" local space-time. Dark matter is therefore visible matter in "its" local universe. Even the Einstein (Misner, 1973) view of gravitation may set-up the link between parallel space-times. In fact curvature in a given space-time dimension may be observed from the different parallel universes that share the same dimension, the totality of the parallel universes are comprised in what we now consider the whole universe or, simply, the universe. If the whole universe contains a superposition of a number of different space-times that are electro-magnetically orthogonal, i.e. space-times in which electromagnetic interaction is confined in its own space-time, then there is more "room" for different forms of matter, which might be reciprocally "incompatible" if mixed directly. Antimatter could be confined in a parallel universe, and the same may be true for other forms of massive particles. According to measurements of the mass ratio between visible matter and dark matter obtained from the observation of the dynamics of a large sample of galaxies, it is a reasonable hypothesis that baryon matter is about 1/5 of the total mass of the whole matter universe. The dark matter issue is therefore resolved by the possible existence of four/five parallel universes, each with almost the same total mass. Space-times with higher dimensionality are frequently studied in theoretical physics; they appear in the Kaluza-Klein theory, string theory and brane theories. In most theories, the additional dimensions are curled to sizes of the order of the Planck length and therefore inaccessible for direct observation. Brane theories (Germani, 2002) predict large extra dimensions and experiments to observe the effects of the possible existence of "macroscopic" additional dimension have been performed with negative result so far. To ensure agreement with these data and to keep a full agreement with the well-known Special Relativity, the proposed model changes our view of reality by giving to "time" the secondary role of derived coordinate. The overall number of fundamental large dimensions is still equal to the observed four, which have now the properties of spatial dimensions.

## DEDUCTION OF THE MODEL

The proposed model was first inferred by attempting to write the wavefunction for gravitons in space-times in terms of a cosmological constant. Imposing the unitarity (the property that defines the probability to find the particle in space-time to be one) of the wavefunction under all possible energy conditions leads to inferring the possible existence of a hyperspace with 4 space-like dimensions, thus, it can be named 4-space. The possibility of a varying cosmological constant (Modanese, 2004), which brings the graviton back to a 3+1 space-time, opens the choice between 4 space-times with 3 space-like dimensions and one of time, therefore four 3+1 space-times should exist. We certainly live in one of these 3+1 space-times; hypotheses can be formulated for the role and the content of the remaining three space-times, and experiments will possibly discriminate among them. To independently show that four 3+1 space-times exist, we need to understand what a 3+1 space-time is and, more specifically, what is the difference between "space" and "time". Unfortunately this is not an easy task. In General Relativity (Misner, 1973) the difference between space and time is encoded in a + or – sign in the metric. It is up to us to define the problem.

It is due to Special Relativity that the difference originates between space and time. The difference is related to the constancy of the speed of light and its invariance for all observers (the Lorentz invariance), which is an experimental fact. It is also an experimental fact that space dimensions are seen to be three, and to this number theories must adapt. It is the scope of this paper to propose that reality can be different. If the difference between space and time is modelled by Lorentz invariance, a different but equivalent formulation of Special Relativity may give some clue. In fact the concept of treating space and time the same way is the scope of Euclidean Special Relativity (ESR). The metric of flat space-time in ESR has the signature ++++, therefore our 3+1 space-time is described embedded in the 4-space. With a closer look to ESR the four 3+1 space-times emerge with ease, and recognizing this possibility is one of the contributions given by this article.

## EUCLIDEAN SPECIAL RELATIVITY

Historically, Euclidean Special Relativity has been first proposed by Montanus (2001) and then further developed by Gersten (2003). Almeida (2001) has recently and independently developed the same concepts and derived a theory named 4-dimensional optics (4DO). Gersten and Almeida recognized that the theory is an extension of ray and wave optics. Recently, Almeida (2004b) has made the interesting attempt to construct a 4DO theory of gravitation. Special Relativity (SR) has been developed to mathematically describe the observation that the speed of light is the same for all observers. This fact led to the dismissal of Galilean transformations in favour of the Lorentz transformation. It became obvious that space and time were both part of a single entity named space-time, of which SR predicted the relevant properties of time dilatation and mass increase. These effects are confirmed by many experiments. Despite its success, SR is often affected by ambiguities of interpretation of the results. In SR, the quantity:

$$(d\tau)^2 = c^2(dt)^2 - (dx)^2 - (dy)^2 - (dz)^2 \tag{1}$$

with $c$ equal to the speed of light, is invariant with respect to a Lorentz transformation. Minkowski proposed considering $t$, $x$, $y$, and $z$ as the coordinates of space-time, $\tau$ being a measure of the distance, named proper time. Equation (1) is the metric of flat space-time; it is a "local" description of the properties of space-time in absence of gravity. In ESR equation (1) is rewritten and discussed:

$$c^2(dt)^2 = (d\tau)^2 + (dx)^2 + (dy)^2 + (dz)^2 \tag{2}$$

In equation (2) $\tau$, $x$, $y$ and $z$ are the coordinates of the 4-space and $t$ is the parameter used to evaluate velocity and acceleration; substantially equation (2) is the definition of time. Time $t$ is an integral local function of changes of the four space coordinates, dimensionally aligned by $c$. Time exists if there is motion in the 4-space. In ESR the measurability of time $t$, that is a function of $\tau$, $x$, $y$ and $z$, implies the existence of an interaction between the measuring instrument and the environment. The Lorentz transformation has been found to be described by SO(4) rotations (Gersten, 2003).

In the present interpretation of ESR the quantities that appear in the right member of equation (2) are homogeneous and have the dimension of space coordinates. With $\tau$ a space coordinate, it is certainly possible to explore this coordinate by some technical means. To discover them, we observe that in Euclidean flat space there is an invariant, the square of the four-velocity, obtained from equation (2) by dividing by $(dt)^2$:

$$\left(\frac{d\tau}{dt}\right)^2 + \left(\frac{dx}{dt}\right)^2 + \left(\frac{dy}{dt}\right)^2 + \left(\frac{dz}{dt}\right)^2 = c^2 \tag{3}$$

In ESR there is a single 4-space in which all particles, for which Lorentz transformation is applicable, travel at the speed of light. Gersten indeed recognized that equation (3) is a restriction in the 4-space, i.e. it defines a subspace in which Lorentz transformation may occur for one particle state. Taking the Lagrangian for a free object of ESR (Montanus, 2001):

$$L = m\left[\left(\frac{d\tau}{dt}\right)^2 + \left(\frac{dx}{dt}\right)^2 + \left(\frac{dy}{dt}\right)^2 + \left(\frac{dz}{dt}\right)^2\right] = mc^2 \tag{4}$$

being $L=K-U$, kinetic energy minus potential energy, we have for a free massive particle $E=mc^2$. The expression is simple because in ESR (in absence of gravity) mass does not depend on velocity and velocity is a universal constant: $E$ is a constant of motion. Since $E \propto m$, $E$ can be easily interpreted to be the energy required to literally "launch" a particle into existence. For particles with velocity in the 3-space equal to zero, it is the first term in left member of equation (3) that is equal to the square of the speed of light. The same applies to the other three possibilities in which particles travel at low speed (compared to $c$) in the 3-spaces $(\tau, x, y)$, $(\tau, x, z)$ and $(\tau, y, z)$. It is only necessary to show that photons travel only in a 3-space to conclude that four electromagnetically orthogonal (i.e. reciprocally invisible for observers at rest in each one) sets of particles (local universes for which Lorentz invariance is applicable) might exist in the 4-space. The existence of four orthogonal sets of particles, which we can consider hosted by four space-times, is motivated by the principle of symmetry: all space dimensions have identical properties, exactly as equation (2) suggests. Local universes travel at the speed of light along the 4 orthogonal coordinates of a reference system for which "directions" and "velocity" are univocally defined. Inside each local universe SR maintains its validity.

Photons travel only in 3-space because for photons in the local universe, for instance in (x, y, z), we have:

$$\left(\frac{dx}{dt}\right)^2 + \left(\frac{dy}{dt}\right)^2 + \left(\frac{dz}{dt}\right)^2 = c^2 \qquad (5)$$

that represents the constancy of the speed of light. Therefore considering the same photons in the 4-space we have:

$$\left(\frac{d\tau}{dt}\right)^2 = 0 \qquad (6)$$

These photons cannot move along the space dimension $\tau$, while the respective matter universe runs away along $\tau$ with speed $c$. The same apply to the remaining three space-times. This is obviously in agreement with SR in (x, y, z, t), in fact photons are particles that have an "onboard clock" completely at rest. When matter increases its speed in a 3-space and approaches the speed of light, it shifts towards another parallel universe, as shown by equation (3). A spacecraft travelling in a 3-space at about half the speed of light should be capable of detecting one or more different parallel universes and in principle should be capable of reaching them. It seems also quite natural to attribute the same behavior to the stellar objects we observe at the edge of our visible universe, objects that seem to fall towards an undefined border.

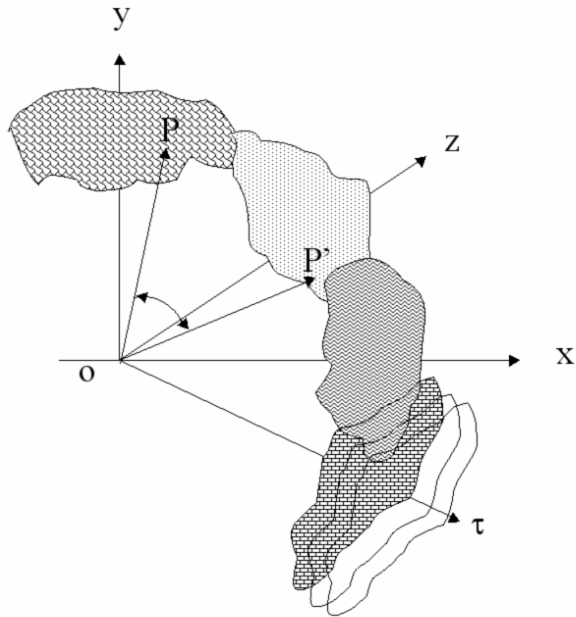

**FIGURE 1.** Local Representation of the Four 3+1 Space-times.

Figure (1) represents the four space-times in a locally Euclidean 4-space. Particles that are at rest (in their own space-time) travel at speed $c$ along their "proper time" coordinate of the 4-space. Our space-time is symbolically made with bricks and consists of an infinite number (a continuum) of $\tau$-frames, some are shown as contour lines. Particle P can reach a different space-time (P->P') with the help of a Lorentz transformation, a SO4 rotation. Traveling between different $\tau$-frames requires the change of the local speed of light, this in turn requires a refractive index different from unity for the propagation of massive particles, and the effect is produced by the presence of a gravitational field.

ESR has its own paradoxes, which are different from those of SR. For instance, if I do the action of switching on a lamp, equation (6) shows that the photons emitted by the lamp in my 3-space will never reach my eyes. In fact my speed along the space dimension $\tau$ is $c$, and the speed of the emitted photons along the space dimension $\tau$ is zero. Therefore the photons do indeed travel in 3-space, towards me, at speed $c$ but they will reach my "ghost" in a past $\tau$-frame. The solution is that the photons that reach my eyes are emitted in some future $\tau$-frame, and my action of switching on the lamp only "seems" to be purely deliberate in my present $\tau$-frame. The 4-space encodes past, present and future in its four dimensions.

For sentient beings, as we are, all information is carried by photons, this fact explains why we can only act in our present. We are a part of a universe that "travels" in space towards the future along the $\tau$ coordinate, our activity in the present "processes" the incoming future and produces the past along the $\tau$ coordinate. The fact that we see photons coming from future $\tau$-frames and from different distances in 3-space, teach us that there is an infinite number of $\tau$-frames before and after our frame and each frame is the present of another material existence. The $\tau$-frames are all alive the same way: past and future $\tau$-frames are as real as our present in the 4-space, therefore it is the 4-space that contains the information, matter is only a process. The philosophical impact of this interpretation of ESR will be discussed later.

Causality is not a property of the 4-space, in which there is no time coordinate. Causality is a property of non-intersecting time-lines (sequences of $\tau$ frames), that are non-intersecting trajectories followed by particles for which Lorentz invariance is applicable. In this picture, local universes follow straight trajectories, free of time-line loops.

## PARTICLES, FORCES AND GRAVITATION IN THE 4-SPACE

ESR has been derived from SR, therefore it inherits a partial experimental validation from SR. 4DO, that is a theory of gravity in 4-space, inherits a partial experimental validation from GR. In fact it is possible to convert the metrics of GR into refractive indices and both theories can give equivalent or comparable prediction (Almeida, 2004b; Montanus, 2001). Obviously, the interpretation of the results is different. New experiments will allow discriminating among the different approaches. According to Almeida (2001), in Euclidean space it is possible to write:

$$c^2(dt)^2 = g_{\alpha\beta} dx^\alpha dx^\beta \tag{7}$$

it has been already shown that in ESR the metric tensor is:

$$g_{\alpha\beta} = \delta_{\alpha\beta} \tag{8}$$

Starting with the study of photon propagation from equations (5) and (6) we get:

$$g_{0\alpha} = g_{\alpha 0} = 0, \quad g_{ab} = n^2 \delta_{ab} \;(a,b \neq 0) \tag{9}$$

with $n$ the refractive index, function of coordinates ($n=1$ without gravitation). Then:

$$c^2(dt)^2 = n^2\left[(dx)^2 + (dy)^2 + (dz)^2\right] \tag{10}$$

introducing the variational principle:

$$\delta \int n d\sigma = 0 \tag{11}$$

with

$$d\sigma = \sqrt{(dx)^2 + (dy)^2 + (dz)^2} \tag{12}$$

it produces the Fermat's principle of optical propagation. Therefore it can be concluded that photons travel as waves with speed $|c/n|$ and wavelength $\lambda=h/(|n|E)$. More generally, extending the study to the 4-space, equations (7), (8) and (11) give:

$$c^2(dt)^2 = n^2\left[(d\tau)^2 + (dx)^2 + (dy)^2 + (dz)^2\right], \quad n \neq 0 \qquad (13)$$

after switching to spherical coordinates, with $d\varphi=0$ and $d\theta=0$, writing $ds^2=n(d\tau^2)$ and substituting:

$$\frac{1}{n} = \left(1 - \frac{2Gm}{r}\right) \qquad (14)$$

the Schwarzschild's metric of General Relativity can be obtained (Almeida, 2001). A comparable model has been proposed by Montanus (2001), employing exponential refractive indices, which agree with GR using a series expansion. It is the possibility of making these links between 4DO and GR that gives 4DO a chance to satisfy various test of GR, those have already passed the experimental verification. The interpretation of the experimental results is obviously different. Future experiments might challenge the two theories.

The refractive index therefore describes the properties of the 4-space, on the other hand the substitution $ds^2=n(d\tau^2)$ indicates that with the presence of matter and assuming GR fully valid, the refractive index is not isotropic, with components that depend on the four directions, this is better shown by resorting to a coordinate transformation (d'Inverno, 1996) with a rigorous calculation of the two refractive indices (Almeida, 2004a). The relations between GR and 4DO indicate that gravity propagates in four dimensions, this obviously holds for weak gravitational waves, which travel at speed c (far from a gravitational field) in the 4-space. The refractive index of the 4-space appears as a square in the Euclidean metric of equation (13) and with $|n|<1$, i.e. for $Gm>r$, the speed in the 4-space is higher than the speed of light. In fact comparing equation (2) with equation (13) the speed of all particles in the 4-space with a refractive index, i.e. with gravity, is $v=c/|n|$. In this situation the Euclidean theory is easier to interpret than General Relativity. The solution for trajectories in the 4-space can be obtained by applying the Fermat's principle in four dimensions, the name 4DO originates from this approach.

Particles are described by waves in the 4-space, because all particles are in motion at some finite speed. Almeida (2001) has suggested identifying these waves with the matter waves of Quantum Mechanics. Equation (7) admits the solution in forms of waves parametrized along $t$:

$$\dot{\Phi}^2 = g^{\alpha\beta}\partial_\alpha\partial_\beta\Phi = \delta^{\alpha\beta}\Phi \qquad (15)$$

Therefore waves appear as soon as a direction is defined, and the process is obviously a "readout" in 4-space. In fact, if the refractive index of the 4-space encodes a pattern, then the motion of the wave/particle reproduces the encoded information. On the other hand the "readout" influences the 4-space as in the example of equation (14), where the mass of a particle associated to a wave changes the refractive index of the 4-space.

By combining $v=c/|n|$ with equation (14) we have:

$$v = c\left|\left(1 - \frac{2Gm}{r}\right)\right| \qquad (16)$$

For small masses and large distances, the presence of a mass reduces the local speed of light, which according to $E=mc^2$ and assuming conservation of energy, changes the local mass of an additional particle in the nearby 4-space to $m(c^2/v^2)$. Therefore mass may diverge if mass density goes above a given limit and energy is conserved. Without going into further details it seems that 4DO is non-linear like GR and therefore capable of offering many interesting developments. Non-linear mass summation can be tracked back to a positive reaction in the refractive index in spite of the fact that the local universe travels at speed c along its fourth dimension, in fact gravity propagates in four dimensions and $n$ exhibits no delay. Photons travel in three dimensions and a propagation loop between two photons at different $\tau$ in four dimensions is not possible: electromagnetism is linear.

ESR and the notion of a refractive index of the 4-space are a significant departure from the concepts of GR: 4-space is flat, it has a refractive index, everything that is not a refractive index is in motion and the velocity is a function of the refractive index. Forces are associated to changes in the trajectory of particles due to the refractive index. The concept of force in a 3+1 space-time as projection of trajectories in the 4-space suggests a simple and unified approach to the problem. Different forces, those that obey to different rules in the 3+1 space-time, are naturally associated to different particles and different properties of the refractive index. If a particle responds to more than one force, it must be a composite particle. With full analogy with optical materials, the refractive index of the 4-space could be a function of frequency, thus allowing a 4DO model for all interactions and explaining the different intensity of the different known forces. Frequency modulation of a particle wavefunction (by vibrations at relativistic speeds in the 3-space) should change its frequency spectrum and control the intensity of natural forces. Experiments very similar to those devised for studying High Frequency Gravitational Waves (HFGW) (Fontana, 2004 and 2003b; Baker, 2004) are suggested to study the properties of the refractive index of the 4-space.

There are speculations about the possibility that the universe might be a gigantic simulation (Bostrom, 2003). After familiarizing with ESR and 4DO, one might be tempted to identify the refractive index with a memory device in which wave/particles play the role of signals flowing through the 4D optical circuits of the universe.

## PERSPECTIVES FOR SPACE TRAVEL AND PROPULSION

The Euclidean view of dimensionality is sufficient to offer a new perspective to space travel. Interactions that are based on electromagnetism are confined in the 3-space and all known propulsion techniques, except one, relate to electromagnetic interaction, often at the molecular level. With 3 degrees of freedom, a large amount of energy is required to give kinetic energy to an object in the 3-space. On the contrary, gravitational propulsion techniques can directly operate in the 4-space, and a well-known example of gravitational propulsion is "gravity assist". In the 4-space all particles travel at the speed of light, therefore the capacity of traveling can be defined as the capacity of steering in the 4-space that is the capacity of making Lorentz transformations. In fact for "gravity assist" no propulsion is required, only the capacity of steering. HFGW have been studied in GR and can be used for "artificial gravity assist" (Fontana, 2000 and 2003; Baker, 2004), similar techniques might be entirely developed within ESR and 4DO. Space travel between different τ-frames seems also possible by locally changing the refractive index, and the effect is produced by the presence of a gravitational field.

The 4-space do not suffer by the metrical distortions of General Relativity that lead to the difficult interpretation and visualization of the results of that theory. In the 4-space, gravitation is encoded in the refractive index and it has been easily demonstrated that inside the event horizon of a Schwarzschild's black hole the 4-velocity of every particle is higher than the speed of light. This is not associated to paradoxes or conceptual difficulties. Unfortunately the counterpart of the Einstein equations for 4DO have not, as of yet, been discovered, except for simple cases derived from GR. It is possible that a combination of properties of the 4-space may contribute to change $n(\tau, x, y, z)$, and these properties should involve particles belonging to the four parallel universes

## CONCLUSION

The properties of the graviton in space-times with a cosmological constant and variants of SR strongly suggest that our 3+1 space-time is a special case in a more fundamental 4-space. The 4-space can host four electromagnetically orthogonal space-times, one of them is our space-time, in which we can only physically operate on a single frame along τ, that is the "present". Gravitational phenomena can allow navigation in the 4-space and, as soon as technology will permit, it will be possible to discover the possible real existence and nature of the remaining three space-times. Observed gravitational phenomena have already shown that the total mass of the universe is four to five times the mass of the visible universe. This observation is encouraging for the presented approach. In addition, recalling the introduction, some parallel universes might be "antimatter universes", therefore they can become a precious source of fuel for spacecrafts equipped with devices or engines capable of producing gravitational fields strong enough to capture even a limited flow of atoms in the parallel universe.

In Euclidean 4-space, extended with the Schwarzschild's solution that suggests the existence of a refractive index, particles can travel with a speed higher than the speed of light without producing paradoxes. Within the simplest model, the most disconcerting property of the 4-space approach is that the perceived "flow of time" is obviously a "readout/modify" process of a predetermined or prerecorded story encoded in the refractive index. The presented "almost static but reactive" view (which leaves the already visited regions of 4-space affected by the process) of the universe differs from the early interpretation of cosmology in GR, which was static and with no "events" or "choices". Abandoning "time $t$" as a principal dimension in favor of four spatial dimensions suggests the interpretation of reality as a "trip" in a prerecorded sequence of possibilities in which material objects follow the laws of physics and sentient beings can act by choosing their paths. It is implicit that the prerecorded sequence can be modified every time a "readout" is performed. Subsequently this happens in the "present", but it may happen in the "future" and in the "past" along the $\tau$ space coordinate respect to our space-time.

Like General Relativity, the presented Euclidean model of reality admits something like time travel. It is not time travel indeed; it is $\tau$ travel. Personal (local) time $t$ cannot be affected because it depends on Lorentz invariance, a $\tau$ traveler can jump into different time-lines (sequences of $\tau$ frames) and follows them actively as soon as the $\tau$ travel machine is switched off. In the 4-space this is indeed space travel.

# ACKNOWLEDGMENTS

The author wishes to thank Robert M. L. Baker jr. for useful discussions.